\documentstyle[12pt,epsfig]{article}
\begin{document}
\baselineskip = 6.5mm
\topmargin= -5mm
 \begin{center}

{\bf Non-equivalence between Heisenberg XXZ spin chain and Thirring model}

\vspace{1cm}

\vspace{1cm}

T. Fujita\footnote{e-mail: fffujita@phys.cst.nihon-u.ac.jp}, 
T. Kobayashi\footnote{e-mail: tkoba@phys.cst.nihon-u.ac.jp}, and 
M. Hiramoto\footnote{e-mail: hiramoto@phys.cst.nihon-u.ac.jp}

Department of Physics, Faculty of Science and Technology, 

Nihon University, Tokyo, Japan

and 

H. Takahashi\footnote{e-mail: htaka@phys.ge.cst.nihon-u.ac.jp}

 Laboratory of Physics, Faculty of Science and Technology, 

Nihon University, Chiba, Japan

\vspace{2cm}

{\Large Abstract}

\end{center}

The Bethe ansatz equations for the spin 1/2 
Heisenberg XXZ spin chain are numerically solved, and 
the energy eigenvalues are determined 
for the anti-ferromagnetic case. We examine the relation between the XXZ 
spin chain and the Thirring model, and show that the spectrum of the XXZ 
spin chain is different from that of the regularized Thirring model.

\vspace{1cm}


\newpage

\section{Introduction}
The symmetry breaking has been discussed quite extensively in varieties 
of field theories  \cite{q1,q2}. It is believed that the spontaneous 
symmetry breaking should accompany a massless boson, and since 
there should not physically exist a massless boson 
in two dimensional filed theory, the symmetry breaking should not 
occur in two dimensional field theory model. 

However, fermion field theory models are quite different in that 
bosons must be dynamically constructed by fermions and antifermions. 
For the fermion field theory models, the Goldstone boson has been 
known only for the current current interaction model 
of Nambu and Jona-Lasinio (NJL model) \cite{q03}. However, recent careful studies 
prove that there is no massless boson in the NJL model after the chiral 
symmetry breaking \cite{q4,q44}. The physics of the chiral symmetry breaking is rather 
simple. The chiral symmetry which is possessed in the NJL Lagrangian 
with massless fermion is broken in the new vacuum since the new 
vacuum is lower than the trivial one. In this case, the $originally$ 
massless fermion acquires the finite mass, and becomes a massive NJL 
model which predicts always a massive boson, and the boson 
can never become massless since the induced fermion mass 
can never be set to zero. 

In the same way, the chiral symmetry breaking 
in the massless Thirring model is broken in the new vacuum, but 
there appears no massless boson \cite{q4,q44}. 

In this Letter, we solve numerically the Bethe ansatz equations for 
the spin 1/2 Heisenberg XXZ model  \cite{q7,q8} and obtain  the energy eigenvalues of 
the antiferromagnetic states. Since  the massless Thirring model 
is  believed to be equivalent to the spin 1/2 Heisenberg XXZ model at the continuum 
limit  \cite{q6}, it should be interesting to compare the results 
of the energy eigenvalues  of the two models at the continuum limit. 

Here, we calculate the  energy eigenvalues of the excited states for  
the spin 1/2 Heisenberg XXZ model and for the regularized Thirring model. 
It turns out that they do not agree with each other. 
The regularized Thirring model has the first excited state with a finite gap 
while the XXZ spin chain has the gapless excitations. 
This indicates that the XXZ spin chain and the regularized Thirring model 
are not equivalent to each other since physical observables are different. 

Here, we discuss the physics behind the difference between the two models. 
The equivalence between the spin 1/2 Heisenberg XYZ model and the massive 
Thirring model is well established  \cite{q6}. But the massless limit in the massive 
Thirring model is a singular point and should not be  taken naively. 
The massless Thirring has two vacua, one which 
corresponds to the trivial vacuum with the chiral symmetry preserved, 
and the other which is a true vacuum with the violation of the chiral symmetry. 
One sees that the true vacuum state 
is lower than the trivial one, and therefore the true vacuum is 
physically realized. From the present analysis, we show  
that the XXZ spin chain cannot be reduced to the Thirring model with the true 
vacuum even though one may mathematically obtain the Thirring Lagrangian 
from the XXZ spin chain. 

The proof for the equivalence  between the XXZ spin chain 
and the Thirring model is based on the naive continuum limit of 
the XXZ model. However, the XXZ model has only one scale, 
and therefore, the physical meaning of the continuum limit is not clear. 
All the physical observables are measured by the lattice constant $a$, and 
thus in order that the $very$ $small$ $a$ makes sense, one should compare $a$ 
with other scale quantity. In this respect, the continuum limit of the XYZ  
can be well defined, but the XXZ model should keep the lattice constant $a$ finite.   
Even if one says that 
one could derive the field theory model at the continuum limit 
( small lattice constant $a$ ), all the observables of this field theory model 
should be measured by the lattice constant $a$. In the Thirring Lagrangian 
derived mathematically from the XXZ spin chain, there is  no scale parameter 
corresponding to the lattice constant $a$, 
and  physically, this indicates that the XXZ spin chain and the Thirring 
model must be different from each other.

This paper is organized in the following way. 
In the next section, we briefly explain the Bethe ansatz solutions 
for the XXZ spin chain. Section 3 treats the massless Thirring model, and 
the energy eigenvalues of the vacuum and the excited states 
with the chiral symmetry breaking 
are discussed. In section 4, we discuss the equivalence between 
the XXZ spin chain and the massless Thirring models. Section 5 summarizes 
what we have learned here.

\section{Heisenberg XXZ model}

Here, we briefly describe the Heisenberg XXZ model. 
The XXZ model has the following Hamiltonian  \cite{q7,q8}
$$ H=J\sum_{i=1}^{N} \left(  S_i^x S_{i+1}^x +  S_i^y S_{i+1}^y +
\Delta S_i^z S_{i+1}^z  \right) \eqno{(2.1)} $$
where $S_i^a$ is a spin operator at the site $i$. $J$ and $\Delta$ denote
the coupling constant and the anisotropy parameter, respectively, 
and $N$ is the site number. 
The periodicity $S_{N+i}=S_i$ is assumed. This Hamiltonian can be numerically 
solved by the exact diagonalization. However, if one wants to discuss 
the excitation spectrum, then one has to have the site number $N$ which 
must be larger than $N=1000$ or so  \cite{q15}. This is practically impossible. 

Fortunately, this model is solved by the Bethe ansatz technique, and  
the Hamiltonian can be diagonalized by the superposition 
of the wave functions  $\phi (z_{n_1},...,z_{n_m}) $ for $m$ down spin case as 
$$ \Psi = \sum_{P} A(n_1,...,n_m) \phi (z_{n_1},...,z_{n_m}) \eqno{(2.2)} $$
where $P$ means all possible permutations of the $n_1,...,n_m$. 
Further, the coefficient $ A(n_1,...,n_m)$ is assumed to be 
of the following shape, 
$$  A(n_1,...,n_m)=\sum_{P_{\mu}}\sum_{P}\exp \left(i\sum_j^mk_{P_j}n_{\mu_j} 
+{1\over 2} \sum_{j<\ell}\varphi_{P_jP_{\ell}}\right)  \eqno{(2.3)} $$
where $k_i$  denote the pseudo-momentum of the down-spin site. 
From the periodic boundary conditions, we obtain the following equations
$$ Nk_j=2\pi \lambda_j +\sum_{\ell} \varphi_{j\ell} \eqno{(2.4)} $$
where $\lambda_i$  are integers running between $0$ and $N-1$ 
with the condition of $\lambda_1 \leq \lambda_2 \leq \  ...\leq \lambda_m$. 
The equation for $\varphi_{j\ell}$ becomes
$$ \cot {\varphi_{j\ell}\over 2}= {\Delta \sin \left({k_j-k_\ell\over 2}\right)\over
{\cos \left({k_j+k_\ell\over 2}\right)-\Delta\cos \left({k_j-k_\ell\over 2}\right) } } 
\eqno{(2.5)}  $$
In this case, we can express the energy eigenvalue $E$ as
$$ E=\left( {1\over 4}N-m \right)\Delta +\sum_{j=1}^m \cos k_j . \eqno{(2.6)} $$
The Bethe ansatz equations (2.5) can be numerically solved by the new iteration 
method which is developed in ref.  \cite{q9,q10}. 

Here, we should write the translation of the coupling constants between the 
spin chain and the Thirring model  \cite{q6}, and the Thirring coupling constant $g$ is 
related to the  $\Delta$ as
$$ g= {4\pi\Delta\over{ 2\pi-\Delta }}.  \eqno{(2.7)} $$
Here, it should be noted that the correspondence between the two models 
is only meaningful for the condition  
$$ \Delta \leq {2\over 5}\pi  \eqno{(2.8)} $$ 
since  $g$ must be smaller than $\pi$  \cite{q6}. 
The vacuum state 
of the field theory corresponds to the state of $S_z=0$, which is 
just the anti-ferromagnetic state. 
In section 4, the numerical results of the excitation spectrum will be discussed. 

Here, we briefly describe the procedure commonly employed to obtain the Thirring model 
Lagrangian from eq.(2.1)  \cite{q14}. By the Jordan-Wigner transformation, one can rewrite 
the Hamiltonian of eq.(2.1) in terms of the spinless lattice fermion 


$$ H = J\sum_{i=1}^{N} \biggl[ {1\over 2}( \psi_i^\dagger \psi_{i+1} +  h.c.) + 
   \Delta \left. \left(\psi_i^\dagger \psi_{i}-{1\over 2}\right) 
\left(\psi_{i+1}^\dagger \psi_{i+1}-{1\over 2}\right)
 \right]  \eqno{(2.9)} $$

This Hamiltonian can be reduced to the massless Thirring 
model Lagrangian below when one takes naively the continuum limit \cite{q14}.

\section{Thirring model}

The massless Thirring model is described by the following 
Lagrangian density  \cite{q11} 
$$  {\cal L} = i \bar \psi  \gamma_{\mu} \partial^{\mu}  \psi 
  -{1\over 2} g j^{\mu} j_{\mu}   \eqno{ (3.1)} $$
where the fermion current $  j_{\mu} $  is given as
$  j_{\mu} = :\bar \psi  \gamma_{\mu} \psi : $. 

This model is studied by the Bogoliubov transformation, and it is found 
that the vacuum has a chiral symmetry broken phase \cite{q4,q44,q3,q5}. 
The vacuum energy 
$ E_{vac}$ as measured from the trivial vacuum ( $ E_{vac}=0 $ ) is given 
$$ E_{vac}=-{L\over{2\pi}}{\Lambda^2\over{\sinh \left({\pi\over g}\right) }} 
e^{-{\pi\over g}} \eqno{(3.2)} $$
where $\Lambda$ and $L$ denote the cutoff momentum and 
box length in this model, respectively, and all of 
the physical quantities must be measured by the $\Lambda$.

Now, in order to compare directly the regularized 
Thirring model prediction to the energy eigenvalues of the XXZ spin chain 
model, we start the equation for the boson in the regularized Thirring model 
where we still keep the box length $L$ finite. The equation for the boson 
wave function becomes just the same as the massive Thirring model \cite{q13} and 
can be written as  \cite{q4,q44}
$$ Ef_{n} = 2E_{p_n}f_{n}
-\frac{g}{L}\sum_{l=-N_0}^{N_0}f_{l}\left(
1+\frac{M^2}{E_{p_{n}}E_{p_{l}}}+\frac{p_{n}p_{l}}{E_{p_{n}}E_{p_{l}}}\right) 
 \eqno{(3.3)} $$
where $p_n$ and $E_{p_n}$ are given as 
$$ p_n = {2\pi\over L}n  \ , \ \ \ \ 
 E_{p_n} = \sqrt{M^2+p_n^2} . \eqno{(3.4)}  $$
Further, the induced fermion mass $M$ is given as the solution of the following 
equation
$$ {g\over L}\sum_{n=-N_0}^{N_0}{1\over{ \sqrt{M^2+p_n^2} }} =1 . \eqno{(3.5)} $$
$N_0$ is related to the cutoff momentum $\Lambda$ as
$$ \Lambda={2\pi\over L} N_0 .  \eqno{(3.6)} $$

In order to connect the present calculation with the spin chain, we write 
the box length $L$ in terms of the lattice spacing constant $a$ as
$$ L=Na  \eqno{(3.7)} $$
with $N=2N_0+1$. Thus, for the large $N$, we obtain 
$$ \Lambda ={\pi\over a}  . \eqno{(3.8)} $$ 

Eq.(3.3) can be easily solved by defining $A$ and $B$ as \cite{q13,q12} 
$$ A=\sum_{n=-N_0}^{N_0}f_n \eqno{(3.9a)} $$
$$ B=\sum_{n=-N_0}^{N_0}{f_n\over{E_n}}. \eqno{(3.9b)} $$
In this case, we obtain $f_n$ as 
$$ f_n={g\over L} {A+{M^2\over{E_n}}B\over{2E_n-E}} . \eqno{(3.10)} $$
Putting $f_n$ into eqs.(3.9), we obtain the eigenvalue equation for $E$, 
and this can be solved in a straightforward manner. 

At this point, we make a comment on the Bethe ansatz solution of 
the massless Thirring model by the technique developed 
by Bergknoff and Thacker \cite{q16}. This method can be applied 
if one introduces the cutoff momentum, and, in this case,
the solution of \cite{q16} becomes quite similar to the solution 
obtained by the massive Thirring model and therefore one obtains the spectrum 
similar to fig. 2 \cite{q9}.  
In this sense, one has to introduce a scale when one wants to describe 
the spectrum of the bosonic states in the Thirring model.

\begin{figure}
\includegraphics*[width=12cm, height=8cm]{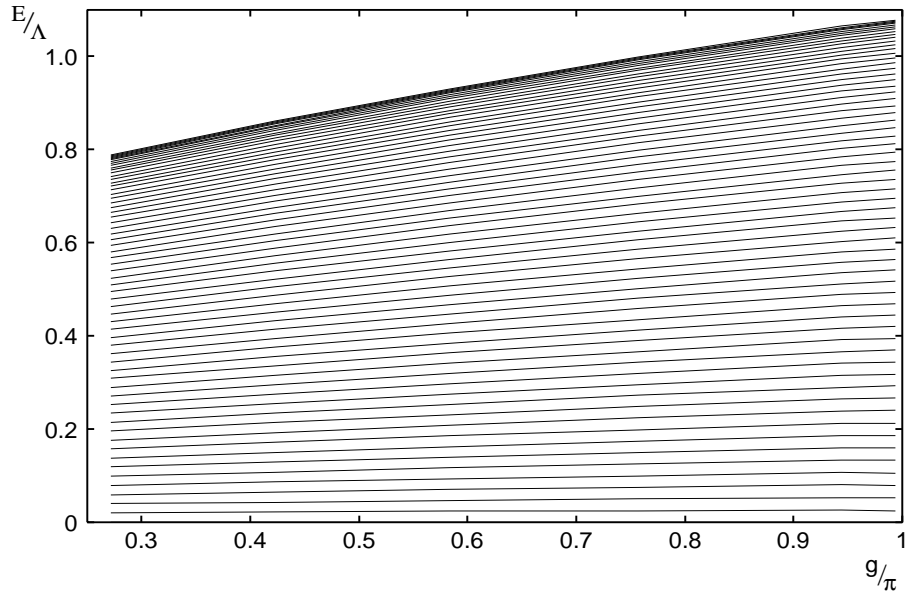}
\caption{The calculated spectrum of the XXZ spin chain}
\end{figure}

\vspace{2cm}
\begin{figure}
\includegraphics*[width=12cm, height=8cm]{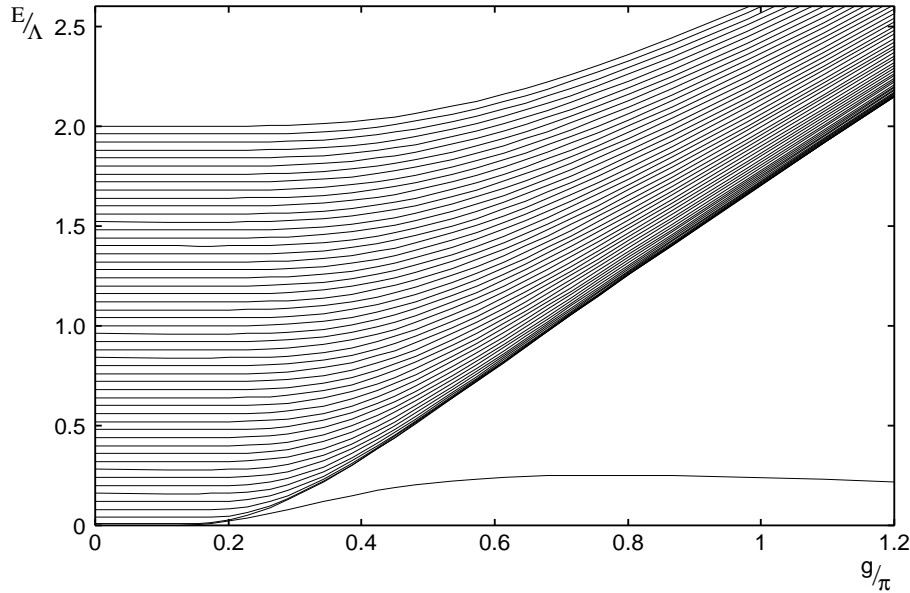}
\caption{The predicted spectrum of the regularized Thirring model}
\end{figure}

\section{Non-equivalence between Heisenberg XXZ \\
and Thirring model}

The Heisenberg XYZ spin chain is known to be equivalent 
to the massive Thirring model at the continuum limit  \cite{q6}. 
The translation of the coupling constants between XYZ and massive Thirring 
is given in eq.(2.7). 

From the above equivalence between the XYZ spin chain and the massive Thirring 
models, one also expects the equivalence between the XXZ and the massless 
Thirring models since the XXZ spin chain corresponds 
to the massless limit of the XYZ spin chain. 
However, the massless limit is a singular point 
in the massive Thirring model, and therefore it is nontrivial 
whether the XXZ spin chain and the massless Thirring model are equivalent 
to each other. In particular, the massless Thirring model has no scale, and 
therefore, one has to introduce the cutoff momentum $\Lambda$ by which 
all of the observables must be measured. On the other hand, the XXZ 
spin chain has a natural scale of the lattice constant, and this is 
an important contrast to the massless Thirring model. 

In figs. 1 and 2, we show the excitation spectrum of the two models. 
Fig. 1 shows the calculated result of the XXZ spin chain while 
fig. 2 shows the predicted spectrum of the regularized Thirring model. 
As can be seen from these figures, there is a significant difference 
between them. In the XXZ spin chain, there is no gap in the first 
excited state, but the regularized Thirring model has a finite gap 
in the first excited state. 
This means that the two models are not equivalent to each other, even though 
it is believed that the XXZ  spin chain at the continuum limit 
corresponds to the massless Thirring model  \cite{q14} if one considers the excitations 
near the fermi sea.  

What is wrong with the derivation of the Thirring model from eq.(2.9) ? 
Here, we present our interpretation of the non-equivalence of the two models. 
In the XYZ spin chain, one can make a continuum limit since there are two parameters 
which have the dimensions, the lattice constant and the mass parameter. 
Therefore, one can make the proper continuum limit in the XYZ spin chain. 
However, when one makes a massless limit from the XYZ to XXZ, then 
the XXZ  possesses only one scale, the lattice constant. In this case, 
one cannot take the continuum limit since everything is already 
measured by the lattice constant. The equivalence between the XXZ spin chain 
and the massless Thirring model derived up to now must be due to the improper 
procedure of the continuum limit in the XXZ spin chain. 
In this respect, one should say that there is no corresponding field 
theory of the XXZ spin chain in the continuum limit, 
and therefore it does not correspond to the massless Thirring model. 

In fact, this continuum field theory of the massless Thirring model 
possesses the chiral symmetry which is not shared by the XXZ spin chain. 
This continuous symmetry plays a very important role for the vacuum 
structure. In the massless Thirring model, 
there are two vacua, one which preserves the chiral symmetry, and 
the other which violates the chiral symmetry. 
Under the chiral symmetry breaking, the true vacuum goes to the one 
which is lower than the trivial vacuum. 
In this respect, the physical 
vacuum of the massless Thirring model is the one that violates the chiral 
symmetry.

\vspace{1cm}

\section{Conclusions}

We have examined the relation between the spin 1/2 Heisenberg XXZ model and 
the massless Thirring model. It turns out that the spectrum of the XXZ model  
is different from the massless Thirring model, and 
it does not possess the true vacuum of the massless Thirring model. 
In this respect, the equivalence between the XXZ and the Thirring models 
does not hold, contrary to the case of the massive theory in which 
the XYZ spin chain is indeed equivalent to the massive Thirring model 
at the continuum limit. 

This is essentially due to the fact that the massless Thirring model 
has the continuous symmetry (chiral symmetry) while the XXZ spin 
chain does not possess such a symmetry. Therefore, in the massless 
Thirring model, there are two vacua, one which keeps the chiral 
symmetry, and the other which violates the chiral symmetry. The true 
vacuum is the one that violates the chiral symmetry since 
it is lower than the other. But the XXZ spin chain cannot reproduce 
the true vacuum state of the massless Thirring model.

\end{document}